\documentclass{amsart}
\usepackage{amssymb}
\usepackage{latexsym}
\usepackage{amsmath}
\usepackage{euscript}
\usepackage{graphics}
\usepackage[all]{xy}

\newcommand{\be}{\begin{equation}}
\newcommand{\ee}{\end{equation}}
\newcommand{\ba}{\begin{eqnarray}}
\newcommand{\ea}{\end{eqnarray}}
\newcommand{\baa}{\begin{eqnarray*}}
\newcommand{\eaa}{\end{eqnarray*}}
\newcommand{\bb}{\begin{thebibliography}}
\newcommand{\eb}{\end{thebibliography}}

\newcommand{\bi}[1]{\bibitem{#1}}
\newcommand{\lab}[1]{\label{#1}}
\newcommand{\re}[1]{(\ref{#1})}

\newcounter{my}
\newcommand{\he}%
   {\stepcounter{equation}\setcounter{my}%
   {\value{equation}}\setcounter{equation}0%
   }%
\newcommand{\she}%
   {\setcounter{equation}{\value{my}}%
    }%

\newcommand\cn{\mbox{cn}}
\newcommand\dn{\mbox{dn}}

\newtheorem{pr}{Proposition}

\theoremstyle{definition}

\numberwithin{equation}{section}

\begin{document}


\title{Birth and death processes and quantum spin chains}

\author{F. Alberto Gr\"unbaum}
\author{Luc Vinet}
\author{Alexei Zhedanov}

\address{Department of Mathematics, University of Califronia, Berkeley, CA 94720, USA}

\address{Centre de recherches math\'ematiques
Universite de Montr\'eal, P.O. Box 6128, Centre-ville Station,
Montr\'eal (Qu\'ebec), H3C 3J7}

\address{Institute for Physics and Engineering\\
R.Luxemburg str. 72 \\
83114 Donetsk, Ukraine \\}

\begin{abstract}
This papers underscores the intimate connection between the quantum walks generated by certain semi-infinite spin chain Hamiltonians and classical birth and death processes. It is observed that transition amplitudes between single excitation states of the spin chains have an expression in terms of orthogonal polynomials which is analogous to the Karlin-McGregor representation formula of the transition probability functions for classes of birth and death processes. As an application, we present a characterization of spin systems for which the probability to return to the point of origin at some time is 1 or almost 1.

\end{abstract}


\maketitle

\section{Introduction}
\setcounter{equation}{0}
This paper aims to point out the relation between birth and death processes and quantum walks generated by (semi-infinite) spin chain Hamiltonians. Birth and death stochastic processes \cite{Feller} are special cases of continuous time random walks (CTRW) on a graph. In CTRWs, the steps on a graph only take place between vertices related by links and occur at time intervals that are sampled from an exponential probability distribution. A CTRW is characterized by a generator matrix with the jumping rates from one vertex to another as entries. The central object of interest are the transition probabilities as a function of time.

The continuous time quantum walk (CTQW) associated to a continuous time random walk is defined  \cite{FG}, \cite{CFG}, \cite{Kemp} through a quantal Hilbert space representation: vertices correspond to state vectors and transitions are governed by the quantum mechanics ensuing from taking the (restriction of the) Hamiltonian on that set of states to be the (Hermitian form) of the generator matrix of the CTRW. 

In this CTQW framework, the key quantities are the transition amplitudes whose unitary time evolutions are thus prescribed by this Hamiltonian. As stressed in \cite{CFG}, the classical transition probabilities and the quantal transition amplitudes are basically related by analytic continuation through the familiar $t \leftrightarrow it $   substitution.

The Markov chains resulting from CTRWs on linear graphs are known as birth and death processes \cite{Feller}. In these cases, the vertices can be labeled by the non-negative integers and transitions only take place between neighboring sites. Birth and death processes have numerous applications, for instance in modeling biological populations and in queuing theory. Their seminal analysis by Karlin and McGregor \cite{KMG1}, \cite{KMG2} rests on the fact that the generator matrix is tridiagonal, allowing for the transition probabilities to be expressed in terms of associated orthogonal polynomials. (For a review see \cite{ILMV}).

Karlin and McGregor adapted the ideas that had been used in the case of diffusion processes by Feller \cite{F1} and McKean \cite{McK}
to study birth and death processes on the nonnegative integers in the cases of discrete \cite{KMc} and continuous \cite{KMG1}, \cite{KMG2}
time parameter. This goes along the spectral theory of self-adjoint operators and orthogonal polynomials play
a crucial role. The earlier work featured a second order differential operator and now we are dealing with a
tridiagonal matrix.

These ideas have been extended in several directions.

In \cite{DRSZ}, \cite{G} one finds two independent developments to deal with the case of quasi birth and death processes,
featuring block tridiagonal matrices and Krein's theory of matrix valued orthogonal polynomials. A special case of this already
arises in \cite{KMc} when the authors attempt to deal with the case of the integers, see the last section of that paper.

A different extension is first considered in \cite{CGMV1} for the case of a discrete time quantum walk either on the integers or the
nonnegative integers. In this case the CMV representation of the unitary matrix U gives the link to (Laurent) orthogonal polynomials on the
unit circle. These ideas have been elaborated in \cite{CGMV2}, \cite{GV},\cite{GMVZ},\cite{GVWW}.

The present paper can be seen as an effort to use the Karlin-McGregor method in the case of quantum walks with continuous time.

In the context of discrete time one knows how to make a quantum walk out of any unitary matrix $U$ given in its CMV form.
On the other hand if one is given an arbitrary Jacobi matrix it is not always possible to make a birth and death process  out of it; extra conditions
on the Jacobi matrix are required. We will see later that a similar situation arises in the case of continuous time parameter. We will
discuss an example of a tridiagonal matrix which fails to give a CTRW but gives rise to a quantum walk.

It may be appropriate to note that the similarities as well as the differences between quantum walks with discrete or continuous time is
an important subject. We refer the reader to the recent paper \cite{Childs} which by  extending
a quantization procedure for discrete time Markov chains due to Szegedy \cite{Szegedi} establishes a correspondence between continuous
and discrete time quantum walks on arbitrary
graphs. In the classical case the use of spectral methods, i.e. Karlin McGregor's formulation is in principle restricted
to reversible Markov chains.

A first point of this article is to underscore the fact the CTQWs corresponding to birth and death processes   can be viewed as single excitation dynamics of spin chains. These systems, among which one has the Heisenberg spin chain, provide fundamental models and are at the heart of significant advances in mathematical physics. They have recently been introduced in the realm of quantum information in the design of (perfect) quantum wires \cite{Bose}.

In particular, it has been shown \cite{Albanese}, \cite{Kay} that with properly engineered couplings, spin chains could effect perfect state transfer, i.e. transport a qubit from one location to another with probability one, without the need of external control.  For such spin chains with couplings that vary from site to site, the hopping of single excitations amounts to CTQWs that parallel birth and death process. The analogy carries mathematically \cite{VZ_PST}, \cite{Chak}   and here again orthogonal polynomials provide a representation of the time dependent transition amplitudes which is similar to the Karlin-McGregor formula for the birth and death process transition probabilities.

A further point of the paper is to identify the conditions for perfect or almost perfect return. This respectively refers to cases where there is probability 1 or almost 1 to return to the initial state of the quantum walk after some time.

The remainder of this paper is organized as follows. In the next section, we review the essentials of the Karlin-McGregor treatment of birth and death processes  in a way that lends itself to the comparison we seek to make. In section 3, we indicate precisely how the CTQWs corresponding to birth and death processes  are connected to the quantum dynamics of spin chains. In Section 4, the natural issue of return is addressed. It is studied using the spectral representation and characteristic functions (Fourier transform) of the weight distribution of the associated orthogonal polynomials. It will be seen that perfect return requires the weight function to be purely discrete on a uniform lattice, while a pure point spectrum ensures almost perfect transfer.

Section 5 provides a number of examples of systems exhibiting perfect return. It is noted in particular that the uniform Heisenberg $XX$ chain does not satisfy the condition for almost perfect return.

\section{The essentials of birth and death processes}
\setcounter{equation}{0}
Birth and death processes are stationary Markov processes whose state space is the set of non-negative integers. For birth and death processes, the transition probability functions form a matrix $P$ whose elements $P_{ij}(t), \: i,j=0,1,2,\dots$ (defined as the probabilities that the system evolves from state $i$ to state $j$ in time $t$)  are required to satisfy the following conditions for infinitesimal time ($t \to 0^{+}$):
\begin{subequations}
  \begin{equation}
    P_{i,i+1}(t) = \lambda_i t + o(t)
    \label{i+1}
  \end{equation}
  \begin{equation}
    P_{i,i}(t) = 1-(\lambda_i + \mu_i)t + o(t)
    \label{i}
  \end{equation}
   \be
   P_{i,i-1}(t) = \mu_i t + o(t) \lab{i-1}
   \ee
\end{subequations}  
Obviously $P_{ij}(0)=\delta_{ij}$ and $P_{ij}(t) = o(t)$ for $|i-j|>1$.  Transitions thus only occur between nearest neighbors. The coefficients $\lambda_i$ and $\mu_i$ assumed to satisfy $\lambda_i>0, \: \mu_{i+1}>0$ for $i \ge 0$ and $\mu_0 \ge 0$ are referred to as the birth and death rates at state $i$, respectively. By hypothesis the conditional probability of the process up to time $t$ depends only on the state of the process at time $t$, this implies that $P_{ij}(s+t)= \sum_{k} P_{ik}(s) P_{kj}(t)$, i.e. $P(s+t)=P(s)P(t)$. This fact, together with \re{i+1}-\re{i-1} readily leads to the (forward) Chapman-Kolmogorov differential equation which is obeyed by the transition probabilities. Indeed one has 
\ba
&& P_{ij}(t + \delta t) = \sum_{k} P_{ik}(t) P_{kj}(\delta t) = \nonumber \\&&P_{i, j-1}(t) P_{j-1,j}(\delta t) + P_{i,j+1}(t) P_{j+1,j}(\delta t) + P_{ij}(t) P_{jj}(\delta t) + O(\delta t) \lab{Ptt} \ea 
from where we get 
\be
\frac{d P_{ij}(t)}{d t} = \lambda_{j-1} P_{i,j-1}(t) + \mu_{j+1} P_{i,j+1}(t) -(\lambda_j + \mu_j) P_{ij}(t) \lab{dP} \ee
In matrix form, \re{dP} can be written 
\be
\frac{dP}{dt} = P(t) A, \lab{matrix_dP} \ee  
where $A$ is the Jacobi (i.e. tri-diagonal) matrix
\be
A=\begin{pmatrix}
 -(\lambda_0 + \mu_0) & \lambda_0 & 0 &  \dots  \\
  \mu_1 & -(\lambda_1+\mu_1) &  \lambda_1 & \dots   \\
   \vdots & & \ddots
   \end{pmatrix} \lab{matrix_A} 
\ee      
which is hence the generator of the CTRW. Note that the Jacobi matrix $A$ is semi-infinite (states $j$ with $j<0$ are forbidden).

In order to solve \re{dP} (or \re{matrix_dP}), following Karlin and McGregor \cite{KMG1}, we introduce the set of orthogonal polynomials $Q_n(x)$ satisfying the 3-term recurrence relation 
\be
-x Q_j(x) = \mu_j Q_{j-1}(x) - (\lambda_j + \mu_j) Q_j(x) + \lambda_j Q_{j-1}(x), \lab{rec_Q} \ee    
with $Q_{-1}=0$ and $Q_0=1$. In compact form
\be
-x Q(x) = A Q. \lab{xAQ} \ee
The spectral parameter $x$ can take either discrete or continuous values depending on a concrete form of the semi-infinite matrix $A$.

It is convenient to introduce the set of polynomials $\{\chi_i(x)\}$ related as follows to the set $\{Q_i(x)\}$
\be
\chi_i(x) = \sum_{j} U_{ij} Q_j(x) \lab{chi_Q} \ee
with 
\be
U_{ij}= (-1)^i \pi_i^{1/2} \: \delta_{ij} \lab{U_ij} \ee
and 
\be
\pi_0=1, \; \pi_i =\frac{\lambda_0 \lambda_1 \dots \lambda_{i-1}}{\mu_1 \mu_2 \dots \mu_i}, \; i>0. \lab{pi} \ee  
They satisfy
\be
x \chi(x) = J \chi(x) \lab{J_chi} \ee
with $J$ the symmetric Jacobi matrix
\be
J= -UA U^{-1} = \begin{pmatrix}
 B_0& J_1 & 0 & 0 & \dots  \\
  J_1 & B_1 & J_2 & 0 & \dots   \\
   0 & J_2 & B_2 & J_3 & \dots \\
\vdots & & & \ddots
   \end{pmatrix} \lab{J_A} \ee
where
\begin{subequations}
\be
B_i=\lambda_i + \mu_i \lab{B_i} \ee
\be
J_i = \sqrt{\lambda_{i-1} \mu_i}. \lab{J_i} \ee
\end{subequations}
From general theorems on the spectral theory of Jacobi matrices \cite{Akhiezer}, there exists a positive measure $d \mu(x)$ \footnote{The support of $d \mu(x)$ can in fact be shown to lie within $[0,\infty)$.} such that the polynomials $\chi_i(x)$ are orthonormal with respect to it:
\be
\int_{-\infty}^{\infty} \chi_i(x) \chi_j(x) d \mu(x) = \delta_{ij}. \lab{ortn_chi} \ee 
Without extra conditions on the coefficients $\lambda_i$ and $\mu_i$ one could have more than one such measure,
   and correspondingly more than one random walk with different boundary conditions at infinity.
   Mathematically this boils down to the possible existence of several self-adjoint extensions of a given symmetric
   operator. Physically this corresponds to the possibility that a particle escapes to infinity in finite time in
   which case one needs to decide what to do at infinity. The reader can consult \cite{Akhiezer} or \cite{AG} for
   a discussion of this issue.

From \re{ortn_chi} it follows that 
\be
\int_{-\infty}^{\infty} Q_i(x) Q_j(x) d \mu(x) = \frac{1}{\pi_i}\delta_{ij}. \lab{ort_Q} \ee
An integral representation for the transition probability functions is now obtained as follows. Introduce the functions
\be
f_i(x,t) = \sum_{j=0}^{\infty} P_{ij}(t) Q_j(x) =[P(t) Q(x)]_i . \lab{f_i} \ee
We have 
\be
\frac{\partial f(x,t)}{\partial t} =\frac{d P(t)}{d t} Q(x) = P(t) A(x) Q(x) = -x f(x,t).\lab{df} \ee
Given that 
\be
f(x,0)=Q(x), \lab{f=Q} \ee
\be
f_i(x,t) = e^{-xt} Q_i(x) \lab{f_i=Q} \ee
and from the orthogonality relation \re{ort_Q} it follows that 
\be
P_{ij}(t) = \pi_i \int_{-\infty}^{\infty} e^{-xt} Q_i(x) Q_j(x) d \mu(x). \lab{P_{ij}} \ee
This is the celebrated Karlin-McGregor integral representation of the transition probabilities. In terms of the polynomials $\chi_i(x)$, it reads 
 \be
 P_{ij}(t) = \left(\frac{\pi_j}{\pi_i} \right)^{1/2} \int_{-\infty}^{\infty} e^{-xt} \chi_i(x) \chi_j(x) d \mu(x). \lab{P_ij_chi} \ee
At this point it is useful to think of a special case. If we had used the set of all integers as our state space,
  and we had chosen $\mu_i=\lambda_i=1$ , so that the matrix above is the discrete Laplacian, the expression for
  $P_{ij}(t)$ would be given by
$\exp(-2t) \mathcal{I}_{i-j}(2t),$
where $\mathcal{I}_k(t)$ stands for the modified Bessel function. Since the eigenfunctions of the discrete Laplacian are given by
the exponentials  $\chi_k(\theta)= \exp(i k \theta)$ the analog of (2.21)  is nothing but the classical expression for $I_k(t)$
in the form \footnote{Here and in the following we trust the reader will not be confused by the occurence of $i$ as label (index) or as the unit imaginary number}
\be
1/(2 \pi)  \int_{0}^{2 \pi} \exp( 2t  \cos \theta ) \exp(i k \theta) d \theta. \lab{Bessel} \ee

\bigskip

In the quantum case , see for instance expression (13) in \cite{CFG} one finds an expression for the transition amplitude between sites $m,l$ in terms of the usual Bessel function $\mathcal{J}_k$. Keeping in mind once again, the difference between the non-negative integers and the set of all integers as a state space, the classical expression for the Bessel functions $\mathcal{J}_k$
is an instance of the expression \re{f_ij_int} below.

In the next section we shall examine how transition amplitudes for certain quantum spin chains compare with this.

\section{Quantum walks from spin chain Hamiltonians}
\setcounter{equation}{0}
The continuous time quantum walk corresponding to the birth and death process characterized by the rate matrix $A$ given in \re{matrix_A} is obtained by assigning vectors $|i\rangle$ to the states $i$ and by using the matrix $J$ of \re{J_A} as the CTQW Hamiltonian. The action of $J$ on the states $|i\rangle$ is naturally taken to be
\be
J |i\rangle = J_{i+1} |i+1\rangle + B_i |i\rangle + J_i |i-1\rangle . \lab{J_basis_i} \ee

A central question is of course to determine the time-dependent transition amplitudes   $f_{ij}(t) = \langle i| e^{iJt}|j\rangle$. It is assumed that the system is finite or semi-infinite, i.e. $i=0,1,2,\dots, N$ for the finite system and $i=0,1,2,\dots$ for the semi-infinite one. In particular, this means that $J_0=0$.

We shall briefly show that this quantum walk can be identified with the single excitation dynamics of a spin chain with inhomogeneous couplings. This will make the connection between birth and death processes and spin chains. Secondly, we shall observe that the amplitudes $f_{ij}(t)$ have an expression which is very similar to that of the probabilities $P_{ij}(t)$ given in \re{P_ij_chi}.

Consider spin chain dynamics governed by Hamiltonians $H$ of the $XX$ type with nearest-neighbor interactions: 
\be
H=\frac{1}{2} \: \sum_{l=0}^{N-1} J_{l+1}(\sigma_l^x \sigma_{l+1}^x + \sigma_l^y \sigma_{l+1}^y) +  \frac{1}{2} \: \sum_{l=0}^N B_l(\sigma_l^z +1), \lab{H_def} \ee 
where $J_l$ are the constants coupling the sites $l-1$ and $l$ and $B_l$ are the strengths of the magnetic field of the sites $l$ ($l=0,1,\dots $). While the number of sites can be finite, we shall here allow for the chain to be possibly semi-infinite. One can adapt the formalism discussed here to the case of a doubly infinite chain following the ideas in \cite{DRSZ} ,\cite{G}. The operator $H$ acts on $\mathbb{C}^{2 \otimes \infty}$ (or $\mathbb{C}^{2 \otimes N}$ in the finite case) with the symbols $\sigma_l^x, \: \sigma_l^y,\: \sigma_l^z$ standing for the Pauli matrices which act as follows on the canonical basis $\{| 0 \rangle, |1 \rangle\}$ for $\mathbb{C}^{2}$ :
\begin{subequations}
\be
\sigma^x | 1 \rangle = | 0 \rangle, \quad \sigma^y | 1 \rangle = i | 0 \rangle, \quad \sigma^z | 1 \rangle = | 1 \rangle,
\lab{sigma_1} \ee
\be
\sigma^x | 0 \rangle = | 1 \rangle, \quad \sigma^y | 0 \rangle = -i | 1 \rangle, \quad \sigma^z | 0 \rangle = -| 0 \rangle .
\lab{sigma_0} \ee
\end{subequations}
The index on these symbols indicate on which factor of $\mathbb{C}^{2}$ they act.

It is immediate to check that 
\be
[H, \frac{1}{2} \: \sum_{l=0}^N (\sigma_l^z +1)]=0
\lab{H_comm} \ee
and it follows that the eigenstates of $H$ split in subspaces labeled by the number of spins over the chain that are in state $| 1 \rangle$.

Let us now restrict $H$ to the eigensubspace spanned by the states with only one excitation, that is, states where all spins but one are in the $|0\rangle$ state . A natural basis for that subspace is given by the vectors 
\be
|i \rangle = (0,0,\dots, 1, 0, 0, \dots), \quad n=0,1,2,\dots, 
\lab{i_basis} \ee  
where the only "1" occupies the $i$-th position. It is straightforward to see that in that basis, the restriction $J$ of $H$ to the one-excitation sector is given by the Jacobi matrix \re{J_A}. Moreover the action on the basis vectors $| i \rangle$ coincide with \re{J_basis_i}. Note that the condition $J_0=0$ is assumed (and amounts to $\lambda_{-1}=0$ in \re{J_i}). This hence establishes the desired identification with the quantum walk stemming from the birth and death process: the quantum states are the single excitation states of the spin chain and the CTQW Hamiltonian $J$ (related to the birth and death processes rate matrix $A$) is simply the restriction of the spin chain Hamiltonian $H$ to these states.

The transition amplitudes are readily seen to satisfy a differential equation of Schr\"odinger type
\be i\frac{d}{d t} f_{ij}(t)= i\frac{d}{d t} \langle i | e^{-iJt} | j\rangle  = \langle i | e^{-iJt} J | j\rangle \lab{dfdt} \ee    
Using a decomposition of the identity, one has 
\be 
i\frac{d}{d t}f_{ij}(t) = \sum_{k=0}^{\infty} \langle i | e^{-iJt} | k \rangle \langle k | J |j \rangle = \sum_{k=0} f_{ik}(t) J_{ki}, \lab{dec} \ee
or in matrix form
\be
i\frac{df }{d t} = f J.  \lab{dfJ} \ee
Equations \re{dec} or \re{dfJ} can now be solved exactly as was done for \re{dP}.

Let 
\be
\Phi_i(t)= \sum_{j}f_{ij}(t) \chi_j(x) = [f(t) \chi(x)]_i . \lab{Phi} \ee 
Clearly,
\be
\Phi_i(x,0) = \chi_i(x). \lab{Phi_0} \ee
We have 
\be
i\frac{d \Phi_i(x,t) }{d t} = \left[i \frac{d f(t)}{dt} \chi(x) \right]_i = [f J \chi]_i= x \Phi_i(x,t). \lab{d_Phi} \ee
Hence
\be
\Phi_i(x,t) = e^{-ixt} \chi_i(x) \lab{Phi_exp} \ee
and using \re{ortn_chi} we find
\be
f_{ij}(t) =  \int_{-\infty}^{\infty} d \mu(x) \chi_i(x) \chi_j(x) e^{-ixt}. \lab{f_ij_int} \ee
In general, for semi-infinite chain, the spectrum (characterized by the measure $d \mu(x)$) can be either discrete or continuous. For the finite chain the spectrum $x_0, x_1, \dots, x_N$  is discrete and finite, in this case the Stieltjes integral in \re{f_ij_int} degenerates into a finite sum over $x_i, i=0, \: \dots, N $. 
We thus observe that, apart from a normalization factor, the transition amplitude $f_{ij}(t)$ has an expression analogous to the Karlin-McGregor formula for the transition probabilities $P_{ij}(t)$ if we make the substitution $t \to it$. (This of course was to be expected given the definition of the CTQW).

As observed after \re{Bessel} an instance of this expression \re{f_ij_int} is given in expression (13) in \cite{CFG}.

Note also that formulas similar to \re{f_ij_int} were obtained earlier in \cite{BM} and \cite{Mantica} in the context of Schr\"odinger quantum dynamics for the lattice Hamiltonians.

\section{Classes of spin chains with perfect and almost perfect return}
\setcounter{equation}{0}
We shall illustrate the usefulness of formula \re{f_ij_int} by considering the problem of perfect and almost perfect return for CTQWs generated by spin chain Hamiltonians of the form \re{H_def}. The probability amplitude for the walker to return to its initial state after some time $T$ is $f_{ii}(T)$. Such a return will be called perfect if $|f_{ii}(T)|=1$. An interesting question is to determine under which conditions can perfect return occur.

Let us first assume that at $t=0$, the chain is in the state $|0 \rangle = (1,0,0,\dots)$ and consider $f_{00}(t)$. From \re{f_ij_int} we have 
\be
f_{00}(t) = \int_{-\infty}^{\infty} e^{-ixt} d \mu(x) =F(-t), \lab{f_F} \ee
where $F(t)$ is the characteristic function of the distribution $d \mu(x)$, defined as the Fourier transform of the measure 
\cite{Lukacs}
\be F(t) = \int_{-\infty}^{\infty} e^{ixt} d \mu(x). \lab{F_def} \ee
The theory of characteristic functions is well developed \cite{Lukacs}. It is known in particular that they possess the properties: (i) $F(0)=1$ and (ii) $|F(t)| \le 1, \; t \ne 0$. In general $|F(t)|<1$ for all $t \ne 0$.  However, there exists a special class of distributions $d \mu(x)$ for which $|F(t_0)|=1$ for some $t_0>0$. This class only comprises (see \cite{Lukacs}) lattice distributions
\be
\mu(x) = \sum_{s=-\infty}^{\infty} M_s \delta(x-\tau_s), \lab{mu_lat} \ee  
where $M_s$ are arbitrary non-negative parameters such that $\sum_{s=-\infty}^{\infty} M_s=1$ and where
\be
\tau_s = \xi + \frac{2 \pi s}{t_0}, \quad s=0, \pm 1, \pm 2, \dots \lab{tau_s} \ee 
with $\xi$ an arbitrary real parameter. 

Recall that the CTQW Hamiltonian is $J$, the one-excitation restriction of the $XX$ spin chain Hamiltonian. We therefore have the following result

\begin{pr} \lab{1}
{\it For a quantum walk initiated in the state  $| 0 \rangle $, there is probability 1 to return to that state after some time $t_0>0$  if and only if  the measure $d \mu(x)$ corresponding to the Jacobi matrix $J$ is a lattice measure \re{mu_lat}.}
\end{pr}

A similar result holds if the initial state is $| i \rangle$ with $i=1,2,\dots$, i.e. if the initial state has its one spin up at the site $i$. In this case the returning amplitude is 
\be
A_{ii} = \int_{-\infty}^{\infty} \chi_i^2(x) e^{-ixt} d \mu(x) =\int_{-\infty}^{\infty} e^{-ixt} d \mu_i(x), \lab{A_nn} \ee  
where
\be
d \mu_i(x) = \chi_i^2(x) d \mu(x). \lab{mu_n} \ee
It is clear that $d \mu_i(x)$ is again a probability measure, because $\chi_i^2(x) \ge 0$ and 
$$
\int_{-\infty}^{\infty} \chi_i^2(x) d \mu(x) = \int_{-\infty}^{\infty} d \mu_i(x)=1.
$$
in view of \re{ortn_chi}. The amplitude $A_{ii}(-t)$ is hence the characteristic function of the distribution $d \mu_i(x)$ and the requirement that $d \mu(x)$ be of the form \re{mu_lat} ensures also that a perfect return to any state $|i \rangle$ ($|A_ii|=1$) is achieved. Note that modified measures of type \re{mu_n} (as well as the corresponding orthogonal polynomials) were studied in \cite{Gau}.

Consider now the generalization to the situation of almost perfect return. This means the following. Assume that the absolute value of the amplitude $f_{00}(t)$ is strictly less then 1 for all $t>0$. Suppose nevertheless that the value 1 can be achieved with any prescribed accuracy for a sufficiently long time $t$. More exactly, this means that there exists a sequence of times $t_n, n=0,1,2,\dots$ such that 
\be
\lim_{t_n \to \infty} |f_{00}(t_n)|=1 \lab{almost_1} \ee    
This is the condition for almost perfect return. 

Again, by a general result from the theory of characteristic functions \cite{Lukacs} it follows that condition \re{almost_1} is equivalent to the condition that the measure $d \mu(x)$ corresponds to a pure point spectrum, i.e. it has the expression \re{mu_lat} but now no other relations like \re{tau_s} are assumed. 

Note that the amplitude $f_{00}(t)$ in this case is an almost periodic function \cite{Lukacs} because it can be represented as a formal Fourier series
\be
f_{00}(t) = \sum_{s=-\infty}^{\infty} M_s e^{-i \tau_s} \lab{f_almost} \ee 
which coincides with one of the possible definitions of almost periodic functions \cite{besic}. We thus have 
\begin{pr}
A quantum walk initiated in the state $|0\rangle$ will almost perfectly return to that state if and only if  the measure $d \mu(x)$ corresponding to the Jacobi matrix $J$ has a pure point spectrum. This is equivalent to the statement that the amplitude $f_{00}(t)$ is an almost periodic function.  
\end{pr}

\section{Examples}
\setcounter{equation}{0}
Let us first offer an example of CTQW corresponding to a linear birth and death process. Take for the birth and death rates the following expressions\
\be
\lambda_i = \frac{c(i+\beta)}{1-c}, \quad \mu_i = \frac{i}{1-c}, \lab{M_lm} \ee   
where $\beta>0$ and $0<c<1$. The corresponding symmetric Jacobi matrix $J$ \re{J_A} has the entries
\be
J_i= \frac{\sqrt{ci(i+\beta-1)}}{1-c}, \quad B_i = \frac{(c+1)i +\beta c}{1-c}. \lab{JB_Meix} \ee

The Karlin-McGregor polynomials $Q_i(x)$ are readily identified with the Meixner polynomials $M_i(x;\beta,c)$ which are orthogonal with respect to the negative binomial distribution \cite{KLS}
\be
\mu(x) = \sum_{s=0}^{\infty} (1-c)^{\beta} \frac{(\beta)_s c^s}{s!} \: \delta(x-s) \lab{Meix_mu} \ee    
(as usual, the symbol $(a)_s =a(a+1) \dots(a+s-1)$ stands for the shifted factorial \cite{KLS}).

This distribution having discrete support, we expect the associated CTQW to exhibit perfect return. In this case, the characteristic function is obtained straightforwardly 
\be
F(t) = \int_{-\infty}^{\infty} e^{-ixt} d\mu(x) = (1-c)^{\beta} \sum_{s=0}^{\infty} e^{-ist} {-\beta \choose s} (-c)^s = \left( \frac{1-c}{1-e^{-it} c} \right)^{\beta}, \lab{char_M} \ee    
where 
$$
{a \choose s} = (-1)^s\frac{(-a)_s}{s!}
$$
is the binomial coefficient. Expression \re{char_M} for the characteristic function of the negative binomial distribution is well known (see, e.g. \cite{Lukacs}). Hence
\be
f_{00}(T) =F(-T)=1 \lab{fF_M} \ee
for $T=\pm 2 \pi n, \; n=0,1,2,\dots$.

Another example with perfect return corresponds to the so-called Stieltjes-Carlitz orthogonal polynomials related to the Jacobi elliptic functions \cite{Chi}.

Consider two symmetric (i.e. with $B_n=0$) orthogonal polynomials $C_n(x)$ and $D_n(x)$ defined by the recurrence relations
\be
J_{n+1} P_{n+1}(x) + J_n P_{n-1}(x) = x P_n(x), \lab{rec_sym} \ee
where
$P_n(x)$ stands for either $C_n(x)$ or $D_n(x)$ and the recurrence coefficients are defined as
\be
J_n^2 = \left\{ k^2 n^2 \quad \mbox{if} \; n \; \mbox{even}  \atop n^2 \quad \mbox{if} \; n \; \mbox{odd}\right . \lab{rec_C} \ee
for the polynomials $C_n(x)$ and 
\be
J_n^2 = \left\{ n^2 \quad \mbox{if} \; n \; \mbox{even}  \atop k^2 n^2 \quad \mbox{if} \; n \; \mbox{odd}\right . \lab{rec_D} \ee
for the polynomials $D_n(x)$. In these formulas $k$ is an arbitrary real parameter satisfying the condition $0<k<1$.

It can be shown that the polynomials $C_n(x)$ are orthogonal on the infinite discrete set of points
\be
\tau_s = \frac{\pi}{2K} (2s+1)m \quad s=0, \pm 1, \pm 2, \dots, \lab{tau_C} \ee
where
\be
K= \int_{0}^{\pi/2} \frac{d \phi}{\sqrt{1-k^2 \sin^2 \phi}} \lab{K_def} \ee
is the complete elliptic integral of the first kind.

These polynomials satisfy the orthogonality relation 
\be
\sum_{s=-\infty}^{\infty} M_s C_n(\tau_s) C_m(\tau_s) = \delta_{nm} \lab{ort_C} \ee
with 
\be
M_s = \frac{\eta_C}{q^{s+1/2} + q^{-s-1/2}}, \lab{M_C} \ee
where $q=\exp(-\pi K'/K)$ is a standard parameter in the theory of Jacobi elliptic functions and $\eta_C$ is an appropriate normalization constant.

Similarly, the polynomials $D_n(x)$ are orthogonal on the infinite discrete set of points
\be
\tau_s = \frac{\pi s }{K}, \quad s=0, \pm 1, \pm 2, \dots \lab{tau_D} \ee    
and satisfy the orthogonality relation 
\be
\sum_{s=-\infty}^{\infty} M_s D_n(\tau_s) D_m(\tau_s) = \delta_{nm} \lab{ort_D} \ee
with 
\be
M_s = \frac{\eta_D}{q^{s} + q^{-s}}. \lab{M_D} \ee
In both cases the spectrum is discrete and linear with respect to $s$. Hence both polynomials $C_n(x)$ and $D_n(x)$ entail condition of the perfect return. Note however that in contrast to the Karlin-McGregor theory, the orthogonality interval is the whole real axis $[-\infty, \infty]$. This means that the recurrence coefficients of the Stieltjes-Carlitz polynomials $C_n(x), D_n(x)$ do not meet the  conditions needed in order to define a Markov birth and death process. For the quantum walks, however, such a restriction on the orthogonality domain is not needed and we have here a perfectly valid example with a double-infinite spectrum. This difference between the classical and quantum walks is worth noting.

It is easily seen that the expressions
\be
f_{00}(t) = \sum_{s=-\infty}^{\infty} M_s e^{-it \tau_s} \lab{f_Fourier} \ee
coincide with the Fourier series of the elliptic Jacobi functions $\cn(z;k)$ and $\dn(z;k)$ \cite{WW}:
\be
f_{00}(t) = \cn(\omega t;k) \lab{f_C} \ee 
for the polynomials $C_n(x)$ and 
\be
f_{00}(t) = \dn(\omega t;k) \lab{f_D} \ee 
for the polynomials $D_n(x)$
with some parameter $\omega$. We see that indeed, both functions are periodic and achieve values 1, as should be for the perfect return. There is an interesting difference between these two amplitudes. Indeed, the amplitude \re{f_C} can take zero values for some values of the time $t$, 
while the amplitude \re{f_D} never achieves zero values on the real axis $t$. This means that the amplitude \re{f_C} possesses the property of "self-avoiding" (i.e. $f_{00}(t_i)=0$ on a uniform lattice $t_i$).

Spin chains with a finite number of sites will have a discrete spectrum and hence will always exhibit almost perfect return (PST).  Among such chains there is a special class which demonstrates so-called perfect state transfer \cite{Albanese}, \cite{VZ_PST}.
This means that for some time $T$ condition $|f_{0N}(T)|=1$ holds. I.e. the state prepared at the site $|0\rangle$ at $t=0$ will reproduce perfectly at the site $|N \rangle$ when $t=T$. Then, by time reversal symmetry, when $t=2T$ this state will perfectly return to the site $| 0 \rangle$.  This means that such chains with PST will always manifest perfect return. However, not every chain with perfect state return will be the chain with PST. Indeed, as we showed, all finite chains demonstrate perfect return, while for PST some restrictions upon the spectrum $x_s$ are needed \cite{Kay}, \cite{VZ_PST}.

Note that for finite spin chains one can introduce a less restrictive almost perfect state transfer condition \cite{Burgarth}, \cite{Godsil}. This means that (for an appropriate time $t$) the difference $1-|f_{0N}(t)|$ can be as small as required with any prescribed accuracy.  Necessary and sufficient conditions for such chains were obtained in \cite{VZ_almost}. It is clear that any spin chain with almost perfect state transfer will demonstrate almost perfect return as well. The inverse statement is not valid.

As we already know, for almost perfect return it is necessary and sufficient that the orthogonality measure $d \mu(x)$ be discrete. In all other cases almost perfect return is impossible. In this connection, consider a semi-infinite uniform Heisenberg $XX$ chain without external magnetic field; this means that $B_i=0$ and that interaction constants are all the same $J_i=1/2,\;  i=1,2,\dots$. The orthogonal polynomials $\chi_n(x)$ corresponding to the Jacobi matrix $J$ coincide in this case with the Chebyshev polynomials of the second kind:
\be \chi_n(x) = \frac{\sin \theta (n+1) }{\sin \theta}, \quad x= \cos \theta .\lab{cheb} \ee 
It is well known \cite{Chi} that these Chebyshev polynomials are orthogonal with respect to the purely continuous measure  
\be
\int_{-1}^1 \chi_n(x) \chi_m(x) \sqrt{1-x^2} dx = \frac{\pi}{2} \delta_{nm} . \lab{ort_cheb} \ee
We thus see that for the uniform Heisenberg chain even almost perfect transfer is impossible.  

This can be confirmed by direct calculation. Indeed, we have
\be
f_{00}(t) =\frac{2}{\pi}\: \int_{-1}^1 \sqrt{1-x^2} e^{-i xt} dx =\frac{2 \mathcal{J}_1(t)}{t}, \lab{f_cheb} \ee  
where $\mathcal{J}_1(t)$ is the first Bessel function of the first kind. The function $f_{00}(t)$ is real, takes value 1 at $t=0$ and oscillates with a decreasing amplitude for $t>0$. The first minimum of $f_{00}(t)$ occurs for $t_1 \approx 5.14$. Its absolute value is $|f_{00}(t_1)| \approx 0.13$. All further local extrema $t_2,t_3, \dots$ correspond to smaller values of $|f_{00}(t)|$. This means that the homogeneous Heisenberg chain is far from showing almost perfect return.

\section{Conclusion}
\setcounter{equation}{0}
We trust this communication is providing an interesting bridge between birth and death processes and quantum walks generated by $XX$ spin chains. The relation between CTRWs on graph and CTQWs is well established in general. We wished to draw the parallel further in the special case where the graph is linear and the generator matrix tri-diagonal, that is when we have a birth and death process. In this instance, the CTQW can be viewed as the single-excitation dynamics of spin chains with nearest-neighbor interaction. Furthermore, as shown by Karlin and McGregor in their analysis of birth and death processes, the theory of orthogonal polynomials offers powerful tools when Jacobi matrices are at play. We here have given indications on how this approach carries in the quantum framework. The study of birth and death processes is a classical topic, we hope that the correspondence with CTQWs will prove helpful in the transfer of results to the quantum walk realm. The spectral results have been put to use to characterize the systems for which the probability to return to the starting point is 1 or almost 1. Almost perfect return occurs if the 1-excitations have a pure point spectrum while perfectness is achieved if the spectrum is uniform in addition.

\bigskip\bigskip
{\Large\bf Acknowledgments}
\bigskip

F.A.G and A.Z. thank the Centre de Recherches Math\'ematiques (CRM) of the 
for its hospitality hospitality while this work was underway.  The work of L.V. is supported in part by a
research grant from the Natural Sciences and Engineering Research
Council (NSERC) of Canada. F.A. Gr\"{u}nbaum acknowledges partial support from the Applied Math.Sciences subprogram of the Office of Energy Research, USDOE, under Contract DE-AC03-76SF00098.

\newpage

\bb{99}

\bi{Akhiezer} N.I.Akhiezer, {\it The classical moment problem and some related questions in analysis}, transl. from the Russian by N.Kemmer, Hafner Publishing Co., New York 1965.

\bi{AG} N.I.Akhiezer and I.M. Glazman, {\it Theory of Linear Operators in Hilbert Space}, 2-nd ed., Dover, 1993.

\bi{Albanese} C.Albanese, M.Christandl, N.Datta, A.Ekert, {\it Mirror inversion of quantum states in linear registers}, Phys. Rev. Lett. {\bf 93} (2004), 230502.

\bi{besic} A.S.Besicovitch  {\it Almost periodic functions}, (Dover, 1954)

\bi{BM} D.Bessis and G.Mantica, {\it Orthogonal polynomials associated to almost
periodic Schrijdinger operators. A trend
towards random orthogonal polynomials}, J. Comp. Appl. Math. {\bf 48} ( 1993) 17--32.

\bi{Bose} S.Bose, {\it Quantum communication through spin chain dynamics: an introductory overview}, Contemporary Physics, {\bf 48} (2007), 13--30.

\bi{Burgarth} D.Burgarth, {\it Quantum State Transfer with Spin Chains}, PhD Thesis (2006); arXiv:0704.1309.

\bibitem{CGMV1}
 M.J.~Cantero, F.A.~Gr\"{u}nbaum, L.~Moral, L.~Vel\'azquez,
 Matrix valued Szeg\H{o} polynomials and quantum random walks,
 Commun. Pure Appl. Math. \textbf{58}, 464--507 (2010).

\bibitem{CGMV2}
 M.J.~Cantero, F.A.~Gr\"{u}nbaum, L.~Moral, L.~Vel\'azquez,
 One-dimensional quantum walks with one defect,
 Rev. Math. Phys. \textbf{24}, 1250002 (2012).

\bi{Chak} R.Chakrabarti, J.Van der Jeugt, {\it Quantum communication through a spin chain with interaction determined by a Jacobi matrix}, J.Phys.: Math Theor. {\bf 43} (2010), 085302.

\bi{Chi} T. Chihara, {\it An Introduction to Orthogonal
Polynomials}, Gordon and Breach, NY, 1978.

\bi{Childs} A.Childs, {On the relationship between continuous- and discrete-time quantum walk}, Comm. Math. Phys.
Volume 294, Number 2 (2010), 581-603

\bi{CFG} A.Childs, E.Farhi and S.Gutmann, {\it An Example of the Difference Between Quantum and Classical Random Walks}, Quantum Information Processing, {\bf 1} (2002), 35--43.

\bibitem{DRSZ} H. Dette, B. Reuther, W. Studden, and M. Zygmunt, {\em
Matrix measures and random walks with a block tridiagonal transition
matrix}, SIAM J. Matrix Anal. Applic. {\bf 29}, No.~1 (2006), 
117--142.

\bi{FG} E.Farhi and S.Gutmann, {\it Quantum computation and decision trees}, Phys.Rev. A {\bf 58} (1998), 915--928.

\bibitem{F1}  W. Feller, {\em On second order differential operators},
Ann. of Math. {\bf 61}, No.~1 (1955), 90--105.

\bi{Feller} W.Feller, {\it An Introduction to Probability Theory and its Applications, Vol. 1,2}, 3-rd Edition, John Wiley and Sons (1968).

\bi{Gau} W.Gautschi and S.Li, {\it A set of orthogonal polynomials induced by a given orthogonal polynomial}, Aequationes Mathematicae, {\bf 46} (1993), 174--198.

\bi{Godsil} C.Godsil, S.Kirkland, S.Severini, J.Smith, {\it Number-theoretic nature of communication in quantum spin chains}, Phys. Rev. Lett. {\bf 109} (2012), 050502; arXiv:1201.4822.

\bibitem{G} F.A. Gr\"unbaum, {\em Random walks and orthogonal polynomials: some challenges},
Probability, Geometry and Integrable Systems, M. Pinsky and B. Birnir ,editors,
MSRI Publication, volume 55 (2007), 241--260, see arXiv math.PR/0703375.

\bibitem{GV}
 F.A.~Gr\"{u}nbaum, L.~Vel\'azquez,
 The quantum walk of F.~Riesz,
 Proceedings of FoCAM 2011, Budapest, Hungary, to be published in the London Mathematical Society lecture Note Series.

\bibitem{GMVZ}
 F.A.~Gr\"{u}nbaum, L.~Vel\'{a}zquez, L.~Vinet, A.~Zhedanov,
 Quantum walks with pure point spectrum,
 in preparation.

\bibitem{GVWW}
 F.A.~Gr\"{u}nbaum, L.~Vel\'azquez, A.~Werner, R.F.~Werner,
 Recurrence for discrete time unitary evolutions,
 arXiv:1202.3903 [quant-ph].

\bi{ILMV} M.E.H. Ismail, J. Letessier, D.R. Masson and G. Valent {\it Birth and death processes
and orthogonal polynomials}. In : Orthogonal Polynomials : Theory and Practice (ed. P.
Nevai), Kluwer Academic Publishers, Dordrecht, 1990, 229--255.

\bi{Kay} A.Kay, {\it A Review of Perfect State Transfer and its Application as a Constructive Tool}, Int. J. Quantum Inf. {\bf 8} (2010), 641--676; arXiv:0903.4274.

\bi{KMG1} S.Karlin and J.L.McGregor, {\it The differential equations of birth-and-death processes and the Stieltjes moment problems}, Trans.Am.Math.Soc. {\bf 85} (1957), 489--546.

\bi{KMG2} S.Karlin and J.L.McGregor, {\it The classification of birth and death processes}, Trans.Am.Math.Soc. {\bf 86} (1957), 366--400.

\bibitem{KMc}  S. Karlin and J. McGregor, {\em
Random walks}, IIlinois J. Math., \textbf{3} (1959), pp. 66--81.

\bi{Kemp} J.Kempe, {\it Quantum random walks: An introductory overview}, Contemporary Physics {\bf 44} (2003), 307--327.

\bi{KLS} R. Koekoek,P. Lesky, R. Swarttouw, {\it Hypergeometric Orthogonal Polynomials and Their Q-analogues}, Springer-Verlag, 2010.

\bi{Lukacs} E. Lukacs, {Characteristic functions}, 2nd Edition, Griffin, London, 1970.

\bi{Mantica} G.Mantica, {\it Quantum intermittency in almost-periodic lattice systems derived
from their spectral properties}, Physica {\bf D 103} (1997) 576--589.

\bibitem{McK} H.P.McKean, Jr., {\em Elementary solutions for certain parabolic partial differential equations},
Trans. Amer. Math. Soc. {\bf 82} (1956), 519--548.

\bi{Szegedi} M. Szegedy, {\it Quantum speed-up of Markov chain based algorithms}. In: Proc. 45th IEEE Symposium on
Foundations of Computer Science, pp. 32-–41, 2004, available at http://arxiv.org/abs/quant-ph/0401053,
2004.

\bi{VZ_PST} L.Vinet, A.Zhedanov, {\it How to construct spin chains with perfect spin transfer}, Phys.Rev. {\bf A 85} (2012), 012323.

\bi{VZ_almost} L.Vinet, A.Zhedanov, {\it Almost perfect state transfer in quantum spin chains}, Phys. Rev. {\bf A 86}, 052319 (2012); arXiv:1205.4680.

\bi{WW} E.T. Whittaker, G.N. Watson, {\em A Course of Modern
Analysis}, Cambridge, 1927.

\eb

\end{document}